\documentclass[
  ,draft            
  ]
  {aipproc}

\layoutstyle{6x9}

\newcommand{\sun}{\ensuremath{\odot}}%

\begin{document}

\title{A study of the mass loss rates of symbiotic star systems}


\author{K. E. Korreck}{
  address={Harvard Smithsonian Center for Astrophysics, 60 Garden Street, Cambridge, MA 02138}
}

\author{E. Kellogg}{
  address={Harvard Smithsonian Center for Astrophysics, 60 Garden Street, Cambridge, MA 02138}
}

\author{J.L. Sokoloski}{
  address={Columbia University, NYC, NY} 
}

\begin{abstract}
The amount of mass loss in symbiotic systems is investigated, specifically mass loss via the formation of jets in R Aquarii (R Aqr).  The
jets in R Aqr have been observed in the X-ray by Chandra over a four year time period.  The jet changes on times scales of a year and new
outflows have been observed.  Understanding the amount of mass and the
frequency of ejection further constrain the ability of the white dwarf
in the system to accrete enough mass to become a Type 1a supernova
progenitor.  The details of multi-wavelength studies, such as speed, density and spatial extent of the jets will be discussed in order to understand the mass balance in the binary system.  We examine other symbiotic systems to
determine trends in mass loss in this class of objects.
\end{abstract}

\maketitle

\section{Introduction}
Symbiotic systems, a hot compact object usually a white dwarf and a
main sequence red giant,  have gained recent attention as the possible
progenitors to Type 1A supernovas.  To reach a Type 1a SN progenitor
the white dwarf must approach 1.4 M$_{\sun}$, however, most of
the white dwarfs in these symbiotic systems have masses $<$ 1
M$_{\sun}$.  An accretion disk around the white dwarf is normally
invoked as part of the method of mass transfer from the red giant to the white
dwarf.  There is great uncertainly as to the
method of mass transfer and the efficiency of that transfer from the red giant to the white dwarf as well as the amount of mass loss from the
white dwarf in nova events and jets. As of 2000, there were 188
symbiotic systems and at least 15 had extended outflows or jets \citep{bel00}.  R Aquarii is unique as it has a spatially extended and resolved x-ray
jet.  CH Cyg has been shown to have a jet with x-ray emission but at a
much smaller spatial extent \citep{gal04}.  This proceeding delineates the mass
involved in the jet of R Aqr as well as a discussion of 5 other symbiotics with jets.

\section{Observations}
The Chandra X-ray observations of the R Aqr jet are detailed in
\citet{kel01} and \citet{kel06}.  The bright x-ray emission in the
Northeast (NE) jet moved with a tangential motion equivalent to
$\sim$600 km s$^{-1}$ in 3.3 yr time interval between the two Chandra
observations. The SW jet, easily seen in 2000.7 observation is
essentially absent in the 2004.0 observation.  Initially, the mass of
the jet was calculated assuming a cylindrical system.  This
calculation did not take into account the true spatial extent of the
emission.  In order to characterize the mass loss in the jet more
quantitatively, the Chandra observation was re-analyzed on a pixel by
pixel basis to give a tighter estimation of the jet volume hence a
tighter mass estimate for the jet in the x-rays.  A summary of the
Chandra x-ray observation and observation in various wavebands of the
jet is presented in Table \ref{tab:a}.

\begin{table}
\begin{tabular}{lrrrr}
\hline \tablehead{1}{l}{b}{Quantity} & \tablehead{1}{r}{b}{Radio}
 $^{(a)}$& \tablehead{1}{r}{b}{Optical} $^{(b)}$ &
 \tablehead{1}{r}{b}{UV}& \tablehead{1}{r}{b}{X-ray}$^{(c)}$ \\ \hline
 Distance from Central Source& 1-9'' & 1-10'' & 1-4''  & 10''\\ Length
 & 10'' & 6'' & 4'' & 5''\\ Mass (M$_{\sun}$) & 3 $\times$10$^{-5}$ & 5
 $\times$ 10$^{-6}$ & 2.3 $\times$ 10$^{-5}$ $^{(d)}$  & 4 $\times$
 10$^{-8}$ \\ Density (cm$^{-3}$)& 2 $\times$ 10$^{5}$ & 10$^{3}$ (NW)
 10$^{2}$(SE) & 10$^{3}$-10$^{4}$ $^{(e)}$ & 100\\ V (km s$^{-1}$)&
 300 &80-130& 36-235 $^{(f)}$& 380\\ \hline \tablenote{(a) All radio
 values are taken from \citet{dou95}, (b) All Optical values are taken
 from \citet{hol91}, (c) All x-ray quantities are taken or derived
 from \citet{kel01,kel06}, (d) \citet{kaf86}, (e) \citet{mei95}, (f)
 \citet{hol97b}}
\end{tabular}
\caption{R Aqr Jet in Multiple Wavelengths}
\label{tab:a}
\end{table}

The mass loss was calculated using the spatial extent of the jet to
determine the volume and the densities given in the references.  The
total emitting mass in all wavelengths for R Aqr jet is 4.5
$\times$10$^{-5}$ M$_{\sun}$.  The quoted mass loss rate for the red
giant in the system is 2.7 $\times$ 10$^{-7}$ M$_{\sun}$ yr$^{-1}$
\citep{hol86}.  The mass loss rate from the red giant is the upper
limit as the mass must be accreted from the red giant to the disk of
the white dwarf then be transfered to the white dwarf and then through
a yet undefined process be ejected as a jet.  Even if the transfer of
mass from the red giant wind to the white dwarf is 10\% efficient, the
accretion rate would  be on the order of 10$^{-8}$ M$_{\sun}$
yr$^{-1}$.  Assuming only another 10\% efficiency in the jet formation
process the assumed rate of mass ejection as a jet would be 10$^{-9}$
M$_{\sun}$ yr$^{-1}$.  Thus it would take 10$^{4}$ years to build up
the jet in R Aqr.  Since the cooling time of the emitting region is on
the order of a few hundred years \citep{kel06}, this long time scale
is unreasonable.

However, if the jet is clumpy, local density goes up but the overall
emitting volume goes down decreasing the amount of mass lost in the
jet. Another alternative is that the jet propels a small amount of
the material and the shock created when this fast material meets the
ISM is simply illuminating the surrounding material that was produced
by large nova events.

\subsection{Mass Loss Through Jets and Outbursts}
The two spatially closest symbiotic systems, R Aqr and CH Cyg have
x-ray jets.  Five other symbiotics are reported to have jets
observable in other wavelengths and are summarized in Table
\ref{tab:b} although the exact rate and amount of mass loss is unknown
as the jets are transient and not monitored in all wavelengths.  Red
giant mass loss rates through winds in symbiotic systems range from
10$^{-9}$ - 10$^{-6}$ M$_{\sun}$ yr$^{-1}$.  If we again assume that
10\% of the wind is transfered to the white dwarf the accretion rates
are of the order 10$^{-10}$ - 10$^{-7}$ M$_{\sun}$ yr$^{-1}$.   Since
jet formation is also poorly understood, 10\% of the accreted mass on
the white dwarf could go into the jet giving an available jet mass
rate of 10$^{-11}$ - 10$^{-8}$ M$_{\sun}$ yr$^{-1}$.

The red giant winds in Table \ref{tab:b} range from 10$^{-8}$
-10$^{-6}$ M$_{\sun}$yr$^{-1}$, giving accretion rates of
10$^{-9}$-10$^{-7}$M$_{\sun}$yr$^{-1}$.  The jet masses from
observations however, vary from 10$^{-5}$ to as little as 10$^{-11}$
M$_{\sun}$ in StH$\alpha$ 190.  These jet masses imply that the rate at
which the mass of the jet is ejected must be higher than expected.

In CV's it is proposed that in order to form a jet the mass accretion
rate had to be between 10$^{-7}$ -10$^{-6}$ M$_{\sun}$
yr$^{-1}$\citep{sok04}.  However, as gleaned from Table \ref{tab:b},
symbiotics form jets, even x-ray jets (strong jets), with lower
accretion rates than the CVs.  As for symbiotics, it seems that jet
formation can occur with accretion rate from the red giant to the
white dwarf as low as 10$^{-8}$ M$_{\sun}$yr$^{-1}$.

\begin{table}
\begin{tabular}{lrrrrr}
\hline \tablehead{1}{l}{b}{Object} & \tablehead{1}{r}{b}{RG Mass Loss}
 &\tablehead{1}{r}{b}{Accretion Rate}& \tablehead{1}{r}{b}{Jet Mass}&
 \tablehead{1}{r}{b}{Compact Object}\\ &\bf{Rate}(M$_{\sun}$/yr) &
 (M$_{\sun}$/yr)&(M$_{\sun}$)&(M$_{\sun}$)\\ \hline  R Aqr &
 2.7$\times$10$^{-7}$  $^{(a)}$ &10$^{-8}$&5$\times$10$^{-5}$ & 0.8
 $^{(b)}$\\  CH Cyg & 2.7$\times$10$^{-6}$  $^{(c)}$ & &
 3$\times$10$^{-7}$  $^{(d)}$ & 0.44 $^{(e)}$\\  Z And & 4$\times$
 10$^{-8}$  $^{(f)}$ & & $<$2 $\times$ 10$^{-7}$  $^{(g)}$ & 0.65
 $^{(h)}$ \\  MWC 560& & 5$\times$10$^{-7}$ & & $>$10$^{-9}$& 0.5
 $^{(i)}$\\ StH$\alpha$ 190 $^{(j)}$& 5 $\times$ 10$^{-8}$&  & 1
 $\times$ 10$^{-11}$& \\ Hen 3-1341$^{(k)}$& & 5$\times$10$^{-8}$&
 &2.5 $\times$10$^{-7}$&0.5\\ \hline
\end{tabular}
\\ \tablenote{a.\citet{hol86}, b.\citet{hol97c},
c. \citet{sko96},d. \citet{gal04}, e. lower limit from \citet{ezu98},
f. \citet{sea90}, g. \citet{bro04} (h) \citet{sch97} (i) All MWC560
values taken from \citet{sch01}, (j) \citet{mun01}, (k)\citet{mun05}}
\caption{Symbiotic Jets}
\label{tab:b}
\end{table}

\section{Conclusions}
Can the white dwarfs in symbiotic systems accrete enough mass to
become Type Ia progenitors?  There is a balance that needs to be
struck with the outflow of material and the influx of material onto
the white dwarf.  Symbiotics have the correct population to be the
progentiors of Type 1a \citep{mun92}.  The determining factor is the
mass accretion rate onto the white dwarf.  Three parameters are
necessary, the red giant wind to white dwarf mass transfer rate, the
disk to white dwarf mass transfer rate and the formation of the jet by
the white dwarf. By careful observations, simulations and modeling of
these parameters, a further constraint on the possibility of growing
symbiotic white dwarfs can be reached.


\begin{theacknowledgments} 
K. E. Korreck would like to thank the American Astronomical Society
for the Travel grant to attend the conference where this work was
presented. The data presented here is from the Chandra X-ray
Observatory Center, which is operated by the Smithsonian Astrophysical
Observatory for an on behalf of the National Aeronautics Space
Administration under contract NAS8-03060.  This work is supported by
the U.S. National Aeronautics and Space Administration, NASA grant,
G04-3050A and contract NAS8-39073.  J. Sokoloski is supported by NSF
Award AST 0302055.  NRAO is supported by Associated Universities,
Inc. under cooperative agreement with the NSF.

\end{theacknowledgments}

\end{document}